\documentclass[10pt,conference]{IEEEtran}
\IEEEoverridecommandlockouts

\usepackage[flushleft]{threeparttable}
\usepackage{cite}
\usepackage{amsmath,amssymb,amsfonts}
\usepackage{algorithmic}
\usepackage{graphicx}
\usepackage{textcomp}
\usepackage[usenames,dvipsnames]{xcolor}
\usepackage{pifont}
\usepackage{mdframed}
\usepackage{url}
 
\usepackage{lipsum,tabularx}

\usepackage{multicol}
\usepackage{multirow}
\usepackage{amssymb}
\usepackage{xcolor}

\usepackage{colortbl}
\usepackage{booktabs}
\usepackage{setspace}

\usepackage[hidelinks,breaklinks,colorlinks]{hyperref}
\hypersetup{
  linkcolor={blue!70!black},
  citecolor={red!70!black},
  urlcolor={blue!70!black}
}
\def\Snospace~{\S{}}

\usepackage[T1]{fontenc}

\usepackage{minted}

\usepackage{balance}

\usepackage{bm}
\usepackage{fp}
\usepackage{siunitx}
\usepackage{amsthm}
\newtheorem{definition}{Definition}

\usepackage[labelfont=bf,font=small,skip=5pt]{caption}
\usepackage{subcaption}

\usepackage{comment}

\usepackage{fancyhdr}
\usepackage{tikz}

\usepackage{xspace}

\usepackage{wrapfig}  
\usepackage{subcaption}
\usepackage{graphicx}

\newcommand{\boxbeg}{
  \noindent\begin{tabular}{|l|}\hline
    \begin{minipage}{3.2in}
      \vspace{2px}
      \noindent
      }

      \newcommand{\boxend}{
      \vspace{2px}
    \end{minipage} \\ \hline
  \end{tabular}
}

\newcounter{finding}
\newcommand{\finding}[1]{\refstepcounter{finding}
    \begin{mdframed}[linecolor=gray,roundcorner=12pt,backgroundcolor=gray!15,linewidth=3pt,innerleftmargin=2pt, leftmargin=0cm,rightmargin=0cm,topline=false,bottomline=false,rightline = false]
    \textbf{Finding \arabic{finding}:} #1
    \end{mdframed}
}

\def\BibTeX{{\rm B\kern-.05em{\sc i\kern-.025em b}\kern-.08em
    T\kern-.1667em\lower.7ex\hbox{E}\kern-.125emX}}

\begin{document}

\title{Where Agent Frameworks Fall Short: Examining Functional Challenges and Usability Concerns}

\author{
\IEEEauthorblockN{
Xinxue Zhu\IEEEauthorrefmark{1}\textsuperscript{*},
Yanzhou Mu\IEEEauthorrefmark{2}\textsuperscript{*},
Xiaoyu Zhang\IEEEauthorrefmark{3},
Tianlin Li\IEEEauthorrefmark{4},
Chao Shen\IEEEauthorrefmark{5},
Chunrong Fang\IEEEauthorrefmark{6},
Yang Liu\IEEEauthorrefmark{3},
and Juan Zhai\IEEEauthorrefmark{7}
}

\IEEEauthorblockA{
\IEEEauthorrefmark{1}Nantong University, China\\
Email: 804134381@qq.com
}

\IEEEauthorblockA{
\IEEEauthorrefmark{2}Ulsan National Institute of Science and Technology (UNIST),
South Korea\\
Email: 602022320006@smail.nju.edu.cn
}

\IEEEauthorblockA{
\IEEEauthorrefmark{3}Nanyang Technological University, Singapore\\
Emails: xiaoyu.zhang@ntu.edu.sg, yangliu@ntu.edu.sg
}

\IEEEauthorblockA{
\IEEEauthorrefmark{4}Beihang University, China\\
Email: tianlin001@buaa.edu.cn
}

\IEEEauthorblockA{
\IEEEauthorrefmark{5}Xi'an Jiaotong University, China\\
Email: chaoshen@mail.xjtu.edu.cn
}

\IEEEauthorblockA{
\IEEEauthorrefmark{6}Nanjing University, China\\
Email: fangchunrong@nju.edu.cn
}

\IEEEauthorblockA{
\IEEEauthorrefmark{7}University of Massachusetts Amherst, USA\\
Email: juanzhai@umass.edu
}

\thanks{\textsuperscript{*}Xinxue Zhu and Yanzhou Mu contributed equally to this work.}
}

\maketitle
\begin{abstract}
Large language model (LLM) agents are increasingly built on agent frameworks that provide reusable abstractions for workflow orchestration, state management, tool integration, and execution control. However, the quality of this infrastructure layer remains insufficiently understood, particularly its functionality challenges and usability concerns, as existing studies have mainly examined traditional deep learning (DL) frameworks or model-level agent failures. Therefore, we conduct an empirical study of 5,669 bug reports and 809 feature requests from five mainstream agent frameworks: \textit{AutoGen}, \textit{CrewAI}, \textit{LangChain}, \textit{LangGraph}, and \textit{MetaGPT}. We construct a four-dimensional taxonomy covering 22 root causes, seven symptoms, 11 motivations, and six requirements, and map them to the five-stage agent lifecycle. 
Across the four RQs, results show an execution centered quality pattern shaped by semantic interface boundaries. Reported bugs mainly manifest as \textit{Incorrect Functionality} (76.00\%) and involve more API, configuration, parsing, and serialization related causes than DL framework bugs, while their associations remain sparse and stage specific. Feature requests mainly target \textit{Feature Enhancement} (49.07\%) and reveal structured needs for \textit{Orchestration Expressiveness}, \textit{Development Delivery}, \textit{Model Adaptation}, and \textit{Tool Ecosystem}. These findings call for quality assurance beyond crash based and tensor level testing, with emphasis on API sequences, structured LLM outputs, serialization boundaries, execution traces, and execution centered maintenance, offering empirical guidance for reliable and usable agent framework infrastructure.\looseness=-1

\end{abstract}

\section{Introduction}
\label{sec:intro}
Large language models (LLMs) have been increasingly used in real-world applications, e.g., intelligent Q\&A~\cite{lewis2020retrieval}, automated software engineering~\cite{chen2021evaluating}, and multimodal autonomous driving~\cite{cui2024multimodal}. As LLM-based applications move toward long-horizon and tool-driven tasks, standalone model invocations are often insufficient, motivating agent systems that integrate LLMs with memory, tools, and control logic. Like mature deep learning (DL) frameworks such as PyTorch~\cite{paszke2019pytorch} support DL-based software, \textit{Agent Frameworks} increasingly serve as the infrastructure layer for agent systems. They provide abstractions for task workflows, agent states, action coordination, and external tool integration. Their quality affects the reliability and maintainability of agent systems, as framework bugs may lead to incorrect task execution, excessive resource use, security risks, or application-level failures. Besides, unclear diagnostics, complex configuration, and limited debugging support can hinder agent development even when no direct functional failure occurs. Therefore, studying both functionality challenges and usability concerns in agent frameworks is necessary for the quality assurance of the broader agent ecosystem.\looseness=-1

Existing studies on agent-related errors can be broadly grouped into two lines of work. The first line investigates functionality challenges in traditional DL frameworks and compilers~\cite{chen2023toward,mu2026deep,zhang2025deep,liu2023nnsmith}, providing systematic taxonomies of bug symptoms and root causes in mainstream DL infrastructures. The second line focuses on failures in agent coordination and orchestration~\cite{xue2025characterization,mu2025understanding,cemri2025multi,wang2024survey}, identifying agent-level challenges such as reasoning errors, hallucinations, biases, and coordination failures through benchmark tasks, controlled experiments, and agent behavior analysis. 
However, these studies still leave several gaps in understanding agent frameworks as an emerging infrastructure layer. 
\textbf{Firstly}, the functionality challenges of agent frameworks themselves remain insufficiently characterized, including their common symptoms and root causes. 
\textbf{Secondly}, developer-facing concerns such as unclear diagnostics, debugging difficulties, configuration friction, and missing framework capabilities have received limited attention, although they directly affect framework adoption, maintenance, and evolution. 
\textbf{Thirdly}, existing studies remain limited on whether findings from DL frameworks generalize to agent frameworks, and where reported issues manifest across the agent lifecycle. 
These gaps motivate a systematic investigation of both functionality challenges and usability concerns in agent frameworks.\looseness=-1

To address these gaps, we conduct a systematic empirical study of bug reports and feature requests in five mainstream agent frameworks: \textit{AutoGen}~\cite{autogenframework}, \textit{CrewAI}~\cite{crewaiframework}, \textit{LangChain}~\cite{langchainframework}, \textit{LangGraph}~\cite{langgraphframework}, and \textit{MetaGPT}~\cite{metagptframework}. Starting from 15,822 raw issue reports collected from their official GitHub repositories, we apply a two-stage filtering pipeline that combines metadata-based screening with manual inspection by five experts, each with at least five years of software engineering experience. This process yields 5,669 bug reports and 809 feature requests as our final dataset. Based on this dataset, we construct a four dimensional taxonomy, covering 22 root cause categories and seven symptom categories for functionality challenges, as well as 11 motivation categories and six requirement categories for usability concerns. We further map each issue to five stages of the agent lifecycle\cite{wang2024survey,li2024survey,dao2026agentic}: \textit{Agent Initialization}, \textit{Perception}, \textit{Self-Action}, \textit{Mutual Interaction}, and \textit{Evolution}. By jointly analyzing reported functional bugs and feature requests among the agent framework communities, our study treats agent frameworks as evolving developer-facing infrastructure rather than only as bug producing systems. This design connects functionality risks with developer-facing improvement needs, complements existing studies on DL infrastructure and agent-level failures, and offers a lifecycle-level perspective on how agent framework issues manifest during development and execution. Overall, our study aims to answer the following research questions (RQs).\looseness=-1

\noindent\(\bullet\)
\textbf{RQ1: How do bug reports in agent frameworks distribute across root causes, symptoms, and lifecycle stages?}
To characterize functionality challenges, we classify bug reports by root causes and symptoms, and map them to lifecycle stages.
The results show that reported bugs concentrate around interface, configuration, parsing, and serialization boundaries, with API related causes accounting for 25.60\% and \textit{Incorrect Functionality} dominating symptoms (76.00\%).
Most reported failures surface in \textit{Self-Action} (92.31\%), indicating an execution centered manifestation pattern rather than a crash centered failure profile.
These findings motivate testing for API sequences, structured outputs, serialization boundaries, state trajectories, and execution traces.\looseness=-1

\noindent\(\bullet\)
\textbf{RQ2: What associations exist among root causes, symptoms, and lifecycle stages in agent frameworks?}
To identify recurring bug patterns, we analyze associations among root causes, symptoms, and lifecycle stages.
The results show a sparse association structure: dominant symptoms concentrate on limited root causes, while frequent API related causes cut across multiple symptoms.
Lifecycle stages show distinct cause and symptom profiles, suggesting that testing and debugging should combine dominant bug patterns with stage specific manifestations.\looseness=-1

\noindent\(\bullet\)
\textbf{RQ3: How do feature requests in agent frameworks distribute across motivations, requirements, and lifecycle stages?}
To understand developer facing improvement needs, we classify feature requests by motivations and requirements, and map them to lifecycle stages.
The results show that requests mainly reflect needs for orchestration expressiveness, development delivery, model adaptation, and tool ecosystem support, while \textit{Feature Enhancement} dominates requirements (49.07\%).
Feature requests mainly target \textit{Self-Action} (64.52\%) but cover more stages than bug reports, suggesting that maintenance should strengthen execution centered capabilities and broader lifecycle support.\looseness=-1

\noindent\(\bullet\)
\textbf{RQ4: What associations exist among motivations, requirements, and lifecycle stages in agent frameworks?}
To identify structured patterns in developer facing requests, we analyze associations among motivations, requirements, and lifecycle stages.
The results show that specific motivations align with limited requirement categories, and that association patterns vary across frameworks and lifecycle contexts.
These findings suggest that maintenance should treat developer requests as structured design signals and align them with framework goals, usage scenarios, and lifecycle contexts.\looseness=-1

Overall, the four RQs show that agent framework quality spans functional failures, developer facing improvement needs, and lifecycle level manifestations. Reported failures expose weaknesses at usage and execution boundaries, while feature requests reveal demands for usable, adaptable, and maintainable infrastructure. These findings support a lifecycle aware quality perspective that connects testing, debugging, maintenance, and framework design.\looseness=-1

Specifically, this work makes four contributions.
\ding{182}\textbf{A Four Dimensional Taxonomy of Agent Framework Issues.}
We construct a taxonomy from 5,669 bug reports and 809 feature requests across five agent frameworks, covering 22 root causes, seven symptoms, 11 motivations, and six requirements. It characterizes reported functionality challenges and developer facing improvement needs.
\ding{183}\textbf{Lifecycle Contextualization of Reported Issues.}
We map bug reports and feature requests to a five stage agent lifecycle. This analysis reveals execution centered reporting patterns and highlights the broad \textit{Self-Action} stage as a target for finer grained analysis.
\ding{184}\textbf{Cross Domain Comparison and Quality Assurance Implications.}
We compare agent framework bugs with DL framework bugs and identify a shift toward API level, configuration related, parsing related, serialization related, and agent specific causes. Based on eight empirical findings, we derive implications for API sequence testing, structured output validation, serialization checking, execution trace analysis, and maintenance planning.
\ding{185}\textbf{Public Dataset and Results.}
We release the dataset and detailed results on our website to support research on reliable, usable, and maintainable agent ecosystems\footnote{https://sites.google.com/view/agent-framework}.\looseness=-1

\section{Background}
\label{sec:background}

\subsection{LLM Agent}
\label{s:bg_agent}

\noindent

An LLM Agent refers to a computational entity that uses LLMs to perceive context, reason over states, and invoke tools or actions to accomplish specific goals
~\cite{xi2025rise,cheng2024exploring}.
It typically possesses properties such as autonomy, reactivity, proactiveness, and social ability~\cite{wooldridge1995intelligent}, allowing it to operate independently while adapting to changes and communicating or collaborating with other agents.
In various domains (e.g., artificial intelligence and distributed systems), LLM agents act as decision-making units that execute complex tasks intelligently, often under uncertain or dynamic conditions~\cite{guo2024large,li2024survey}.\looseness=-1

Researchers have proposed benchmarks and fine-grained harnesses to evaluate agent capabilities~\cite{liu2024agentbench,zhou2024webarena,jimenez2024swebench,qin2024toolllm,ma2024agentboard,shen2024taskbench} and diagnose multi-agent failures via taxonomies and attribution tools~\cite{cemri2025multi,zhang2025agent}.
Additionally, recent work proposes diverse red-teaming and sandbox-based execution to probe agent safety issues~\cite{perez2022red,ganguli2022red,zou2023universal,ruan2024toolemu}.
While these efforts focus on whether agents behave correctly in the applications, this study characterizes the functional errors and usability gaps in the underlying agent frameworks that fundamentally determine whether agents can be built and run reliably.

\subsection{Agent Lifecycle}
\label{s:bg_lifecycle}

\begin{figure}[t] 
\centering 
\includegraphics[scale=0.16]{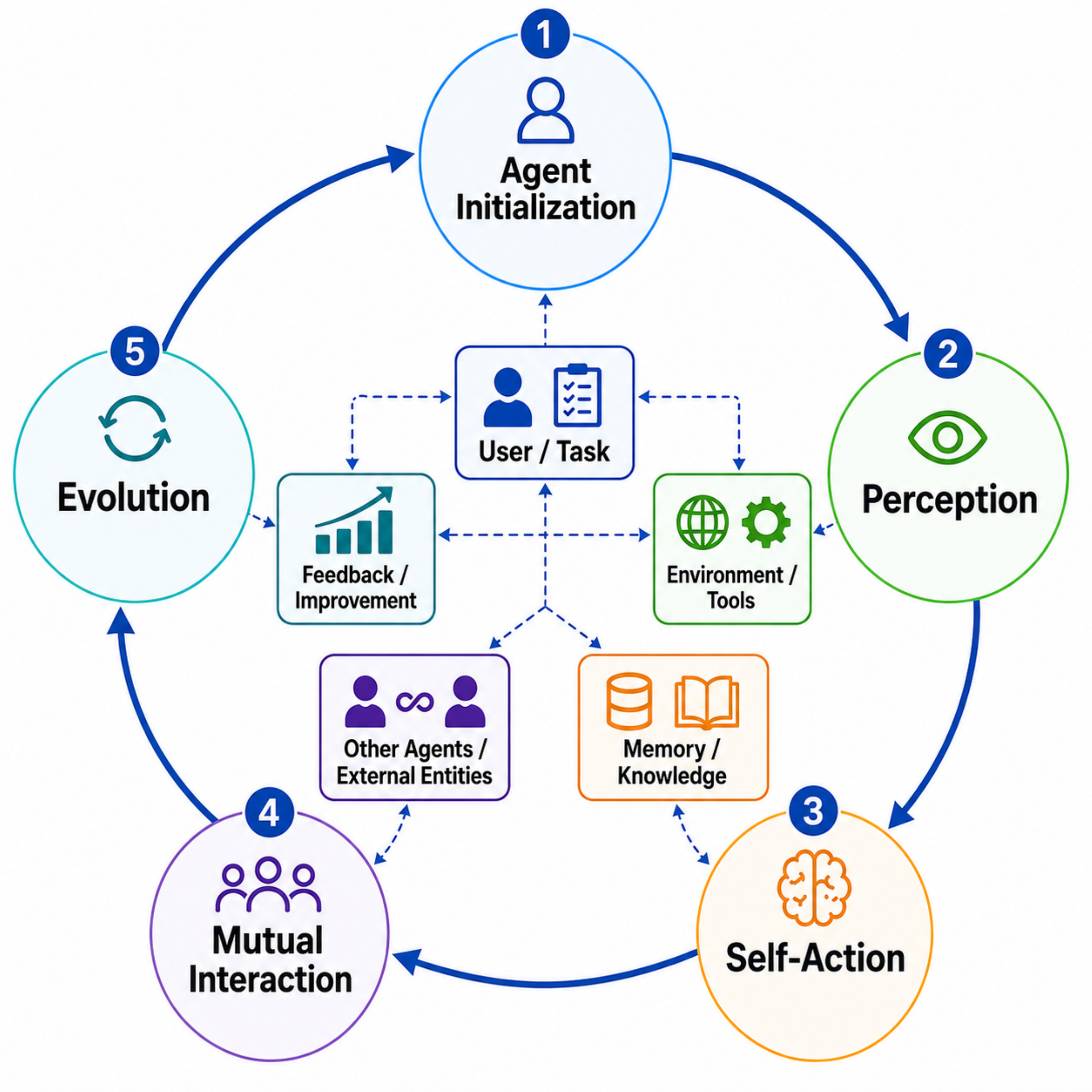} 
\caption{A Five-Stage Lifecycle Model for LLM Agent Systems}
\label{fig:lifecycle} 
\end{figure}

Drawing upon existing studies on LLM agent systems~\cite{wang2024survey,li2024survey,dao2026agentic}, we adopt a five stage lifecycle model to characterize agent construction and execution: \textbf{(1) Agent Initialization}, \textbf{(2) Perception}, \textbf{(3) Self-Action}, \textbf{(4) Mutual Interaction}, and \textbf{(5) Evolution}. \textit{Agent Initialization} specifies identity, role parameters, task objectives, and behavioral constraints. \textit{Perception} handles input ingestion, parsing, and interpretation from users, environments, tools, or peer agents. \textit{Self-Action} captures autonomous execution, where agents reason, make decisions, invoke tools, and complete tasks based on perceived information, memory, and domain knowledge. \textit{Mutual Interaction} covers communication, coordination, state synchronization, and collaboration. \textit{Evolution} supports reflection, feedback incorporation, memory update, and iterative improvement. This lifecycle model, summarized in~\autoref{fig:lifecycle}, provides a structured lens for analyzing where functionality challenges and usability concerns appear across agent frameworks.\looseness=-1

\subsection{Agent Frameworks}
\label{s:bg_framework_scope}

Although existing literature has extensively discussed LLM agents and their architectures, it still lacks a widely accepted operational definition of agent frameworks. Therefore, we define agent frameworks based on development practices and representative open-source projects~\cite{wang2025empirical,derouiche2025agentic}.

\begin{definition}[Agent Framework]
An agent framework is reusable software infrastructure that facilitates the development and execution of LLM-based agent systems.
It provides programming abstractions and runtime support for workflow orchestration, state management, tool integration, and execution control, enabling developers to build, coordinate, and maintain agents with capabilities such as planning, memory, tool use, and action execution.
\end{definition}

From a functional perspective, an agent framework typically supports the development of agents with four core capabilities~\cite{wang2024survey,xi2025rise}.
\ding{182} \textit{Planning} refers to the agent's capacity to decompose complex tasks into manageable subtasks based on user objectives and environmental states, and to formulate execution strategies through reasoning~\cite{wei2022chain,yao2023tree}.
\ding{183} \textit{Memory} refers to mechanisms for storing, maintaining, and retrieving historical interactions, state transitions, and intermediate results, which help agents maintain behavioral coherence across long-horizon workflows~\cite{zhang2025survey}.
\ding{184} \textit{Tool use} denotes the capability to interact with external software, APIs, databases, and other computational resources through interfaces, adapters, or function calls~\cite{schick2023toolformer,qin2024toolllm}.
\ding{185} \textit{Action execution} translates reasoning and planning outcomes into concrete operations, such as invoking tools, updating workflow states, generating outputs, or interacting with external environments~\cite{yao2023react}.

\subsection{Agent Infrastructure Testing}
\label{s:bg_insfra_testing}
Agent infrastructure refers to the software stack that supports the development, execution, and maintenance of AI agents. In this work, we focus on agent frameworks as the agent-specific layer of this infrastructure, while comparing them with adjacent AI infrastructure such as DL frameworks, DL compilers, and LLM optimization libraries~\cite{li2021deep,zhang2025deep,wang2024survey}.\looseness=-1

Prior studies have characterized bugs in DL frameworks~\cite{chen2023toward,mu2025understanding,zhang2025deep,mu2026deep}, tested framework APIs through fuzzing and differential testing~\cite{deng2022fuzzing,guo2020audee,wei2022free,xie2022docter,mu2025devmut,pham2019cradle,wang2020lemon,mu2025improving}, and developed generation-based and coverage-guided fuzzers for DL compilers~\cite{liu2023nnsmith,liu2022tzer}. Recent work also examines reliability and supply chain risks in LLM optimization libraries~\cite{wang2025large,jiang2025foundation}. However, agent frameworks remain less understood as developer-facing infrastructure. Although recent work has characterized functional bugs in LLM agent workflow orchestration frameworks~\cite{xue2025characterization}, it mainly focuses on bug reports and leaves developer requests and lifecycle-level manifestations underexplored. We therefore study agent frameworks from two complementary dimensions: \textit{functionality challenges} and \textit{usability concerns}.

Following prior studies~\cite{zhang2025deep,yusop_revised_2020} and the scope of this work, we use the following operational definitions:

\begin{definition}[Functionality Challenge]
A functionality challenge denotes a reported failure that prevents an agent framework from delivering its intended functional behavior. It may originate from an internal implementation fault, incorrect API use, or a mismatch between the expected runtime configuration and the actual environment. Our classification therefore focuses on the affected functionality and reported triggering condition, rather than assuming that every report corresponds to a framework source code bug.
\end{definition}

\begin{definition}[Usability Concerns]
A usability concern denotes a developer-facing friction or design shortcoming that hinders the use, configuration, integration, debugging, extension, or maintenance of an agent framework, even when it does not directly cause a functional failure.
\end{definition}

We use GitHub feature requests as observable proxies for usability concerns when they contain actionable evidence of user-facing friction, missing support, integration barriers, workflow inefficiencies, or insufficient configurability. Requests without actionable content are excluded during manual inspection.
Our study extends prior agent framework bug studies by jointly analyzing reported functionality challenges and feature requests across five widely used LLM agent frameworks. It constructs a four-dimensional taxonomy covering failure causes, symptoms, request motivations, and requirement types, and maps both failures and requests to a five-stage agent lifecycle. This design connects reliability risks with developer evolution needs and supports implications for semantic interface validation, structured output validation, execution trace checking, and maintenance planning.

\section{Methodology}
\label{sec:methodology}

\begin{figure}[t] 
\centering 
\includegraphics[width=\linewidth]{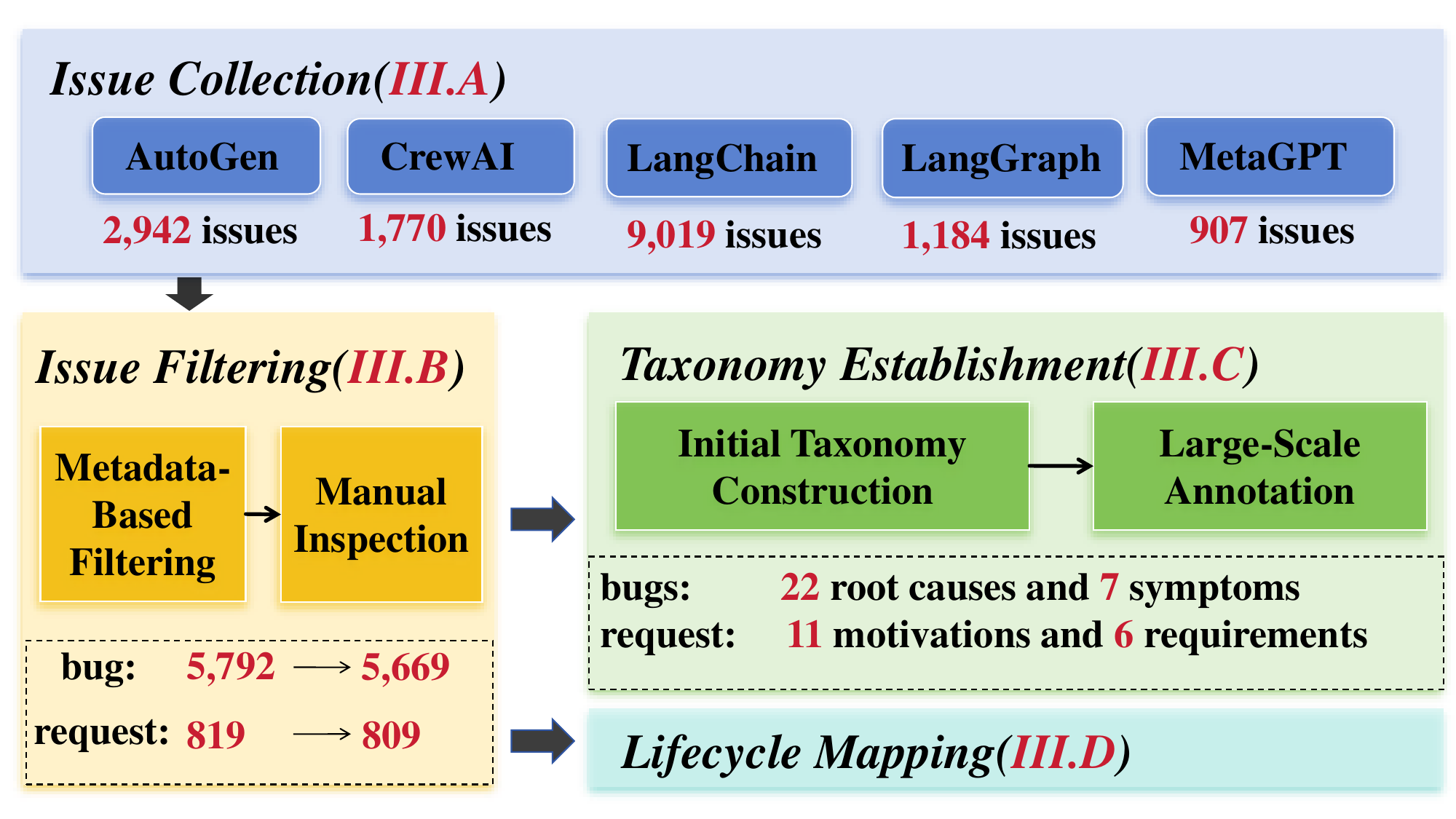} 
\caption{Workflow of Our Four-phase Empirical Methodology}
\label{fig:flowchart} 
\end{figure}

To investigate functionality challenges and developer-facing requests in agent frameworks, we design a four-phase empirical methodology in \autoref{fig:flowchart}. \textit{Issue Collection} selects representative frameworks and collects issue reports from official repositories. \textit{Issue Filtering} applies inclusion and exclusion criteria to retain valid bug reports and feature requests. \textit{Taxonomy Establishment} derives structured categories through iterative annotation and expert discussion. \textit{Lifecycle Mapping} assigns each issue to an agent lifecycle stage using issue evidence and framework APIs.\looseness=-1

\subsection{Issue Collection}

Issue collection determines the empirical scope of our study. To obtain representative agent frameworks, we first compile 99 candidate projects by cross-referencing academic surveys~\cite{liu2024large} and open-source discussions. We then apply our framework definition in \autoref{s:bg_framework_scope} to retain projects that provide agent-specific abstractions for development and execution. We exclude general-purpose SDKs, standalone workflow engines without agent abstractions, end-user applications, and projects that do not target agent systems.

Among the remaining candidates, we rank projects by GitHub Stars as a proxy for community adoption and select the top five frameworks: \textit{AutoGen}~\cite{autogenframework}, \textit{CrewAI}~\cite{crewaiframework}, \textit{LangChain}~\cite{langchainframework}, \textit{LangGraph}~\cite{langgraphframework}, and \textit{MetaGPT}~\cite{metagptframework}. Each selected framework has more than 30.8k stars at the time of collection. These frameworks cover diverse design philosophies, including modular composition in \textit{LangChain}, graph-based orchestration in \textit{LangGraph}, dialogue-centered interaction in \textit{AutoGen}, and role-based collaboration in \textit{CrewAI} and \textit{MetaGPT}. This selection supports both popularity and design diversity, which helps our analysis capture common and framework-specific quality issues.

To avoid bias from partial sampling, we collect all open and closed issues from the official GitHub repositories of these five frameworks between January 2023 and March 2026. This process yields 15,822 raw issue reports, including 9,019 from \textit{LangChain}, 2,942 from \textit{AutoGen}, 1,770 from \textit{CrewAI}, 1,184 from \textit{LangGraph}, and 907 from \textit{MetaGPT}. For each issue, we collect the \textit{title}, \textit{body}, \textit{labels}, and \textit{comments}. These fields provide complementary evidence for later analysis: titles and bodies describe symptoms, contexts, and reproduction steps; labels support preliminary filtering; comments capture developer discussions that often reveal root causes, fixes, or feature motivations.

\subsection{Issue Filtering}

To ensure that taxonomy construction and lifecycle mapping rely on relevant and analyzable reports, we refine the raw issue list through a two-stage filtering process. The first stage uses repository metadata to obtain broad candidates, while the second stage uses expert inspection to remove false positives and low-information cases.

\noindent
\textbf{Stage I: Metadata Filtering.}
This stage uses GitHub labels and issue types that maintainers assign during project maintenance. For functionality challenges, we retain issues whose \texttt{Labels} or \texttt{Types} fields contain ``\texttt{bug}''. For usability concerns, we account for framework-specific labeling conventions and retain issues whose \texttt{Labels} or \texttt{Types} fields contain ``\texttt{feature}'', ``\texttt{enhancement}'', or ``\texttt{feature request}''. After removing unmatched reports, we obtain 5,792 bug candidates and 819 feature candidates.

\noindent
\textbf{Stage II: Manual Inspection.}
Because repository metadata may contain noisy or ambiguous labels, we invite five experts, each with at least five years of software engineering experience, to independently inspect the remaining candidates according to the definitions in \autoref{s:bg_framework_scope}. The experts resolve disagreements through group discussion. For bug candidates, we exclude four types of false positives: \ding{182} \textit{uninformative reports} that lack diagnostic details (e.g., \cite{autogen5643}), \ding{183} \textit{usage questions} that seek task guidance (e.g., \cite{langchain32319}), \ding{184} \textit{cosmetic fixes} unrelated to runtime misbehavior (e.g., \cite{crewAI2378}), and \ding{185} \textit{theoretical security advisories} that lack observed buggy executions (e.g., \cite{langchain10962}). For feature candidates, we exclude reports whose bodies and comments lack actionable content (e.g., \cite{autogen3924}). This filtering process yields 5,669 functionality challenges and 809 usability concerns, which form the final corpus for taxonomy construction and lifecycle mapping.

\subsection{Taxonomy Establishment}

Building on the filtered dataset, this part presents how to construct a four-dimensional taxonomy.

\noindent
\textbf{Stage I: Initial Taxonomy Construction.}
We randomly sample 200 bug reports and 200 feature requests as the seed corpus, and two software engineering experts independently annotate them. For bug reports, we use an established DL framework bug taxonomy~\cite{chen2023toward} as the initial schema for root causes and symptoms, and extend it when agent framework issues cannot be captured by existing categories. For feature requests, the experts derive motivation and requirement categories from the pilot set. They then compare annotations, resolve disagreements through discussion, and refine category definitions until consensus is reached. This process provides validated baseline concepts and supports the comparison between DL and agent framework bugs in~\autoref{sec:rq1}.

\noindent
\textbf{Stage II: Large-Scale Annotation.}
For full annotation, all issues are presented in randomized order, with access to titles, bodies, labels, comments, and linked pull requests when available. Duplicates are merged only when maintainers explicitly mark them as duplicates; otherwise, they are retained as independent reports because they reflect independent developer observations. All five experts independently annotate every issue in the filtered corpus. Each bug report receives one primary root cause and one primary symptom, while each feature request receives one primary motivation and one primary requirement. We use primary labels because GitHub issues may mention multiple contextual factors, but usually emphasize one central failure trigger, symptom, motivation, or requirement. When multiple labels are plausible, experts compare rationales and determine the final label by majority agreement. \looseness=-1

We measure inter-annotator agreement before consensus resolution using Fleiss' kappa~\cite{fleiss1971measuring}. The annotation achieves $\kappa=0.70$ for root causes, $\kappa=0.60$ for symptoms, $\kappa=0.68$ for motivations, and $\kappa=0.74$ for requirements, indicating moderate to substantial agreement. Remaining disagreements are resolved through expert discussion and majority voting. During annotation, experts also flag conflicting cases and emerging patterns for group review. To control taxonomy growth, a new category is added only when \ding{182} no existing category reasonably captures the pattern, \ding{183} more than 10 independent reports exhibit the same characteristic, and \ding{184} the category reflects a conceptually distinct mechanism rather than a minor variant. Through this process, we obtain 22 root causes and seven symptoms for functionality challenges, together with 11 motivations and six requirements for usability concerns.\looseness=-1\looseness=-1

\subsection{Lifecycle Mapping}

To support workflow-level analysis, we map each functionality challenge and usability concern to one of the five agent lifecycle stages defined in~\autoref{s:bg_lifecycle}. The lifecycle label captures where an issue is primarily observed or where a requested improvement is operationally targeted, rather than where the underlying cause is introduced. In particular, \textit{Self-Action} covers a broad execution region, including reasoning, planning, tool invocation, memory access, state processing, and task execution; therefore, a \textit{Self-Action} label should be interpreted as a manifestation or target stage, not as a causal origin.\looseness=-1

The same five experts independently inspect the title, body, and discussion thread of each issue, and assign one primary lifecycle stage. We use single-stage labels to keep the analysis comparable across issues. When an issue involves multiple stages, experts choose the stage where the user observable failure or requested improvement is most directly reflected. To reduce subjective bias, they follow the lifecycle definitions in~\autoref{s:bg_lifecycle} and consult the official API documentation of the five frameworks. Explicit API mentions are used as direct evidence when the documentation links the API to a lifecycle stage; otherwise, mapping relies on issue descriptions, reproduction steps, comments, and lifecycle definitions.

We compute Fleiss' kappa before consensus resolution and obtain $\kappa=0.83$, indicating strong agreement. Remaining disagreements are resolved through expert discussion and majority voting. This process assigns all 5,669 functionality challenges and 809 usability concerns to primary manifestation stage labels for subsequent stage-level analysis.

\section{Result Analysis}
\subsection{RQ1: Agent Framework Failures}
\label{sec:rq1}

\begin{figure}
    \centering
    \includegraphics[width=0.4\textwidth]{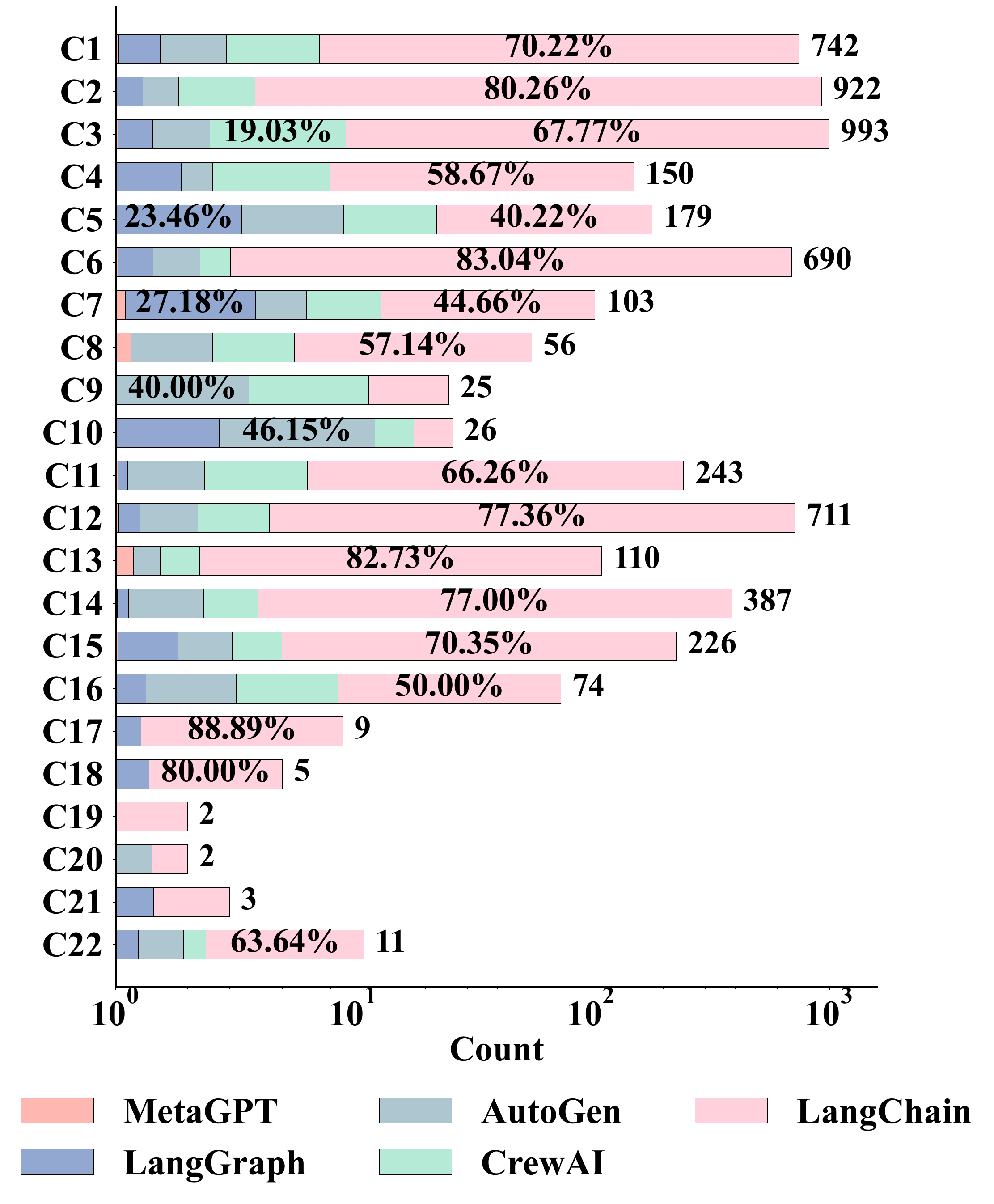}
    \caption{Counts and Agent Framework composition of Root Causes
    }
    \label{fig:rq1rootcause}
\end{figure}

\begin{figure}
    \centering
    \includegraphics[width=0.4\textwidth]{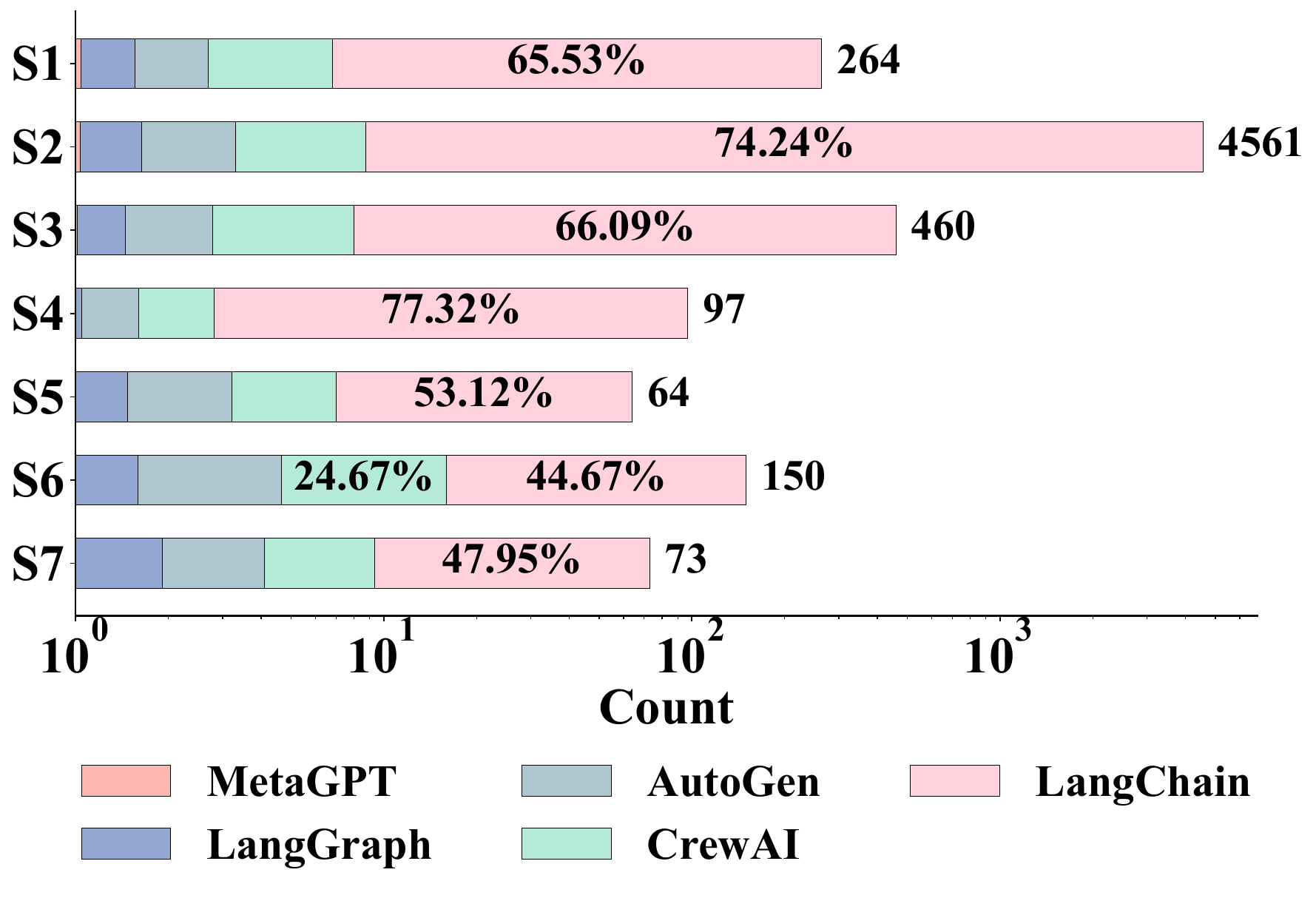}
    \caption{Counts and Agent Framework Composition of Symptoms.
    }
    \label{fig:rq1symptom}
\end{figure}

To understand reported functionality challenges in agent frameworks, we analyze bug reports from three perspectives: lifecycle stage, reported cause, and observable symptom. 
For reported causes and symptoms, we compare agent framework bugs with DL framework bugs from prior work~\cite{chen2023toward}. 
Following the sampling strategy of prior DL framework research~\cite{chen2023toward}, we randomly sample 250 issues from each framework and exclude \textit{MetaGPT} from this comparison because it contains fewer than 250 bug reports. 
We then analyze lifecycle stages using the full filtered bug dataset. \autoref{tab:rq1_root_cause_categories} and \autoref{tab:rq1_symptom_categories} summarize the sampled comparison set, while \autoref{fig:rq1rootcause} and \autoref{fig:rq1symptom} report the full filtered dataset.

\begin{table*}[t]
\centering
\scriptsize
\renewcommand{\arraystretch}{1.15}
\caption{
The Composition of Root Causes.
\scriptsize{
(\textcolor[HTML]{D60000}{Red}/\textcolor[HTML]{2BB73F}{green} marks higher/lower proportions than DL frameworks. Agent percentages are computed on the sampled set.)
}
}
\label{tab:rq1_root_cause_categories}
\begin{tabular}{|p{2.4cm}|p{1.7cm}|p{1.4cm}|p{11cm}|}
\hline
\textbf{Category} & \textbf{Agent framework (\%)} & \textbf{DL framework (\%)} & \textbf{Explanation in Agent Frameworks} \\
\hline

\textit{C1. API Incompatibility} & 12.00\%(\textcolor[HTML]{D60000}{+9.10\%}) & 2.90\% &
This root cause  refers to changes in function signatures, object patterns, or invocation protocols of abstractions that break existing agent configurations and pipelines. \\
\hline

\textit{C2. API Misuse} & 13.60\%(\textcolor[HTML]{D60000}{+1.80\%}) & 11.80\% &
This category refers to functional failures caused by incorrect use or misunderstanding of framework APIs, including execution abstractions, control-flow semantics, tool invocation protocols, and state management. Although the immediate cause may lie in user code, the failure manifests at the framework-usage boundary and prevents the framework from delivering its intended behavior. \\
\hline

\textit{C3. Misconfiguration} & 18.90\%(\textcolor[HTML]{D60000}{+5.50\%}) & 13.40\% &
This category refers to functional failures caused by mismatches between the framework's expected runtime configuration and the actual environment, such as dependencies, credentials, model settings, tool registration, or execution parameters. These issues are classified as framework functional problems because they disrupt the framework's expected functionality under its configuration contract. \\
\hline

\textit{C4. Telemetry Malfunction} & 3.80\% & -- &
This root cause  involves monitoring, tracing, or observability mechanisms that interfere with or fail during execution, interrupting the main workflow due to strict validation, intrusive exception handling, or network certificate issues. \\
\hline

\textit{C5. Concurrency Issue} & 5.70\%(\textcolor[HTML]{D60000}{+2.40\%}) & 3.30\% &
This root cause  manifests as inconsistent or prematurely terminated asynchronous event streams coordinating model outputs, tool responses, or execution signals, or as misunderstandings of the concurrency model.\\
\hline

\textit{C6. Serialization Error} & 10.60\% & -- &
This root cause  refers to faults in encoding or decoding structured execution data.\\
\hline

\textit{C7. Memory Persistence Bug} & 4.00\% & -- &
This root cause  occurs when mechanisms for maintaining long-term agent state, embedding caches, or retrieval indices fail to remain consistent with evolving knowledge sources or model configurations due to design flaws.\\
\hline

\textit{C8. Knowledge Transmission Bug} & 1.30\% & -- &
This root cause  appears in hierarchical or multi-agent architectures as bugs in context transfer logic. \\
\hline

\textit{C9. Console Interaction Bug} & 0.90\% & -- &
This root cause arises when the agent framework’s assumptions about interactive execution environments conflict with the actual runtime context.  \\
\hline

\textit{C10. Frontend--Backend Mismatch} & 1.10\% & -- &
This root cause occurs in frameworks that provide a web interface, where bugs in frontend--backend state synchronization logic cause the frontend to display outdated or inconsistent information.\looseness=-1  \\
\hline

\textit{C11. Documentation Desync} & 4.60\% & -- &
This root cause  reflects inconsistencies between the framework documentation and its actual implementation. \\
\hline

\textit{C12. Model Output Parsing Error} & 9.40\% & -- &
This root cause  arises when content returned by the language model deviates from the expected format, length, or contains invalid characters, preventing the parser from extracting valid information. \\
\hline

\textit{C13. Resource Limitation} & 0.70\% & -- &
This root cause  involves constraints imposed by resource caps, such as token limits, memory, time, or concurrency, or improper resource management. \\
\hline

\textit{C14. Dependent Module Issue} & 6.60\%(\textcolor[HTML]{D60000}{+4.00\%}) & 2.60\% &
This root cause  occurs when external services return errors, time out, or impose rate limits, or when the model itself lacks support for a required feature and the framework fails to handle such scenarios. \\
\hline

\textit{C15. Incorrect Algorithm Implementation} & 4.20\%(\textcolor[HTML]{2BB73F}{-11.00\%}) & 15.20\% &
This root cause  refers to bugs at the implementation level of the framework’s own code. \\
\hline

\textit{C16. Environment Incompatibility} & 2.00\%(\textcolor[HTML]{2BB73F}{-6.60\%}) & 8.60\% &
This root cause refers to errors caused by overlooking specific characteristics of the execution environment, such as hardware platforms or operating systems. \\
\hline

\textit{C17. Incorrect Exception Handling} & 0.10\%(\textcolor[HTML]{2BB73F}{-5.40\%}) & 5.50\% &
The root cause refers to bugs in exception handling, including missing exceptions that should be thrown or handled, redundant exceptions that should not be thrown, and incorrect or imprecise exception messages. \\
\hline

\textit{C18. Type Issue} & 0.10\%(\textcolor[HTML]{2BB73F}{-14.10\%}) & 14.20\% &
This root cause involves type-related problems, such as errors in type conversion or type checking. \\
\hline

\textit{C19. Tensor Shape Mismatch} & 0.00\%(\textcolor[HTML]{2BB73F}{-12.20\%}) & 12.20\% &
This root cause occurs when tensor shapes are incompatible in shape-related operations, such as shape inference or tensor transformation. \\
\hline

\textit{C20. Incorrect Assignment} & 0.00\%(\textcolor[HTML]{2BB73F}{-5.30\%}) & 5.30\% &
This root cause refers to errors caused by incorrectly assigning values to variables or by missing initialization. \\
\hline

\textit{C21. Numerical Issue} & 0.10\%(\textcolor[HTML]{2BB73F}{-3.30\%}) & 3.40\% &
This root cause is caused by incorrect numerical computation, such as division by zero, overflow or underflow, incorrect operators or operands, or missing operands. \\
\hline

\textit{C22. Others} & 0.30\%(\textcolor[HTML]{2BB73F}{-1.30\%}) & 1.60\% &
This root cause encompasses issues that do not fit into any of the previous twenty-one categories. \\
\hline

\end{tabular}
\end{table*}

\noindent \textbf{Analysis of Root Cause Distribution.} 
In the sampled comparison set, reported causes concentrate on configuration and interface issues. 
\textit{Misconfiguration} ranks first (18.90\%), while \textit{API Misuse} (13.60\%) and \textit{API Incompatibility} (12.00\%) together account for 25.60\% of sampled bug reports. 
Compared with DL frameworks, agent frameworks contain significantly more API-level bugs (25.60\% vs. 14.70\%, two proportion $z$ test, $p \le 0.001$), suggesting that many reported functionality challenges arise at the boundaries among framework abstractions, external services, and developer-supplied configurations. 
Agent frameworks expose orchestration-centered interfaces, such as message schemas, tool call contracts, callback protocols, state transitions, and control flow policies;
rapid changes in LLM services, toolkits, vector stores, and runtime environments further increase interface and configuration inconsistencies. 
They also introduce reported cause categories that prior DL framework taxonomies do not separately capture, including \textit{Model Output Parsing Error} (9.40\%) and \textit{Serialization Error} (10.60\%), since LLM generated structured outputs, conversation states, memory objects, and tool results flow across heterogeneous components. 
Malformed \texttt{JSON}, inconsistent schemas, and incompatible serialized states can break execution even when the model call succeeds. 
Framework architecture further shapes reported causes: for example, \textit{AutoGen} reports more \textit{Console Interaction} and \textit{Frontend Backend Mismatch} issues because its dialogue centered multi agent orchestration interacts with both console interfaces and AutoGen Studio.\looseness=-1

\finding{
Agent framework bugs shift from low-level algorithmic and numerical faults toward semantic interface faults. 
This suggests that agent framework testing should prioritize API sequence testing, structured output mutation, schema validation, and serialization boundary checking.\looseness=-1
}

\begin{table*}[t]
\centering
\scriptsize
\renewcommand{\arraystretch}{1.15}
\caption{
The Composition of Symptoms.
\scriptsize{
(\textcolor[HTML]{D60000}{Red} and \textcolor[HTML]{2BB73F}{green} indicate increases and decreases relative to DL frameworks, respectively. Agent framework percentages are based on the sampled comparison set.)
}
}
\label{tab:rq1_symptom_categories}
\begin{tabular}{|p{2.4cm}|p{1.7cm}|p{1.4cm}|p{11cm}|}
\hline
\textbf{Category} & \textbf{Agent framework (\%)} & \textbf{DL framework (\%)} & \textbf{Explanation in Agent Frameworks} \\
\hline

\textit{S1. Crash} & 5.50\%(\textcolor[HTML]{2BB73F}{-45.90\%}) & 51.40\% &
This symptom refers to a visible crash of an agent framework during operation, where the program fails to start, terminates unexpectedly at runtime, or becomes unable to execute its core functionality. \\
\hline

\textit{S2. Incorrect Functionality} & 76.00\%(\textcolor[HTML]{D60000}{+51.70\%}) & 24.30\% &
This symptom refers to observable inconsistencies between the expected behavior and the actual runtime behavior of an agent framework. \\
\hline

\textit{S3. Build Failure} & 9.30\%(\textcolor[HTML]{2BB73F}{-9.50\%}) & 18.80\% &
This symptom refers to situations where an agent framework or application cannot successfully enter a runnable state during deployment, installation, or startup.\\
\hline

\textit{S4. Poor Performance} & 1.50\%(\textcolor[HTML]{2BB73F}{-0.60\%}) & 2.10\% &
This symptom refers to observable declines in runtime efficiency during agent execution. \\
\hline

\textit{S5. Hang} & 1.70\%(\textcolor[HTML]{D60000}{+1.40\%}) & 0.30\% &
This symptom refers to a state in which an agent or execution graph becomes stalled without triggering explicit errors or termination signals. \\
\hline

\textit{S6. Unreported} & 3.70\%(\textcolor[HTML]{2BB73F}{-0.60\%}) & 3.10\% &
This symptom refers to situations where an agent framework accepts inputs or configuration changes but does not react as expected, resulting in unresponsiveness or ignored functionality without explicit error reports. \\
\hline

\textit{S7. Display Anomaly} & 2.30\% & -- &
This symptom refers to observable inconsistencies or errors in user-facing displays within an agent framework, including interface components, visualization panels, log outputs, or documentation pages. \\
\hline

\end{tabular}

\end{table*}

\noindent
\noindent \textbf{Analysis of Symptom Distribution.}
In the sampled comparison set, agent framework bugs mainly manifest as functional result failures with weak crash signals.
\textit{Incorrect Functionality} dominates the symptom distribution (76.00\%), with \textit{Incorrect Execution Result} and \textit{Missing Execution Result} as its main subcategories. 
This pattern indicates that many bugs do not terminate execution, but instead distort task results, omit expected outputs, or violate workflow semantics after reasoning, tool invocation, and state propagation. 
This symptom profile differs from DL frameworks~\cite{chen2023toward}, where prior bugs more often involve tensor computation, operator implementation, compilation, or hardware backend interactions and thus more frequently appear as \textit{Crash} or \textit{Build Failure}.
In contrast, agent frameworks rely on dynamic interaction, probabilistic model outputs, external tools, and long-context workflows, while also incorporating a degree of fault tolerance, allowing execution to continue even when erroneous or incomplete outcomes are produced.
This contrast aligns with higher proportions of \textit{Incorrect Functionality} and \textit{Hang}, and lower proportions of \textit{Crash} and \textit{Build Failure}, in agent frameworks. 
For example, \textit{CrewAI} reports frequent \textit{Hang} symptoms because hierarchical task delegation and inter agent dependencies can introduce repeated coordination, blocking waits, or unresolved workflow states.\looseness=-1

\finding{
Agent framework functionality challenges often take the form of silent semantic failures with weak crash signals.
The decline of \textit{Crash} from 51.40\% in DL frameworks to 5.50\% in agent frameworks suggests that crash-focused test oracles alone cannot effectively detect agent framework bugs.
Agent framework testing therefore needs semantic oracles, such as state trajectory validation, expected output checking, tool call trace inspection, and execution trace replay.
}

\noindent
\noindent \textbf{Analysis of Lifecycle Distribution.}
We map the 5,669 bug reports to the five agent lifecycle stages. 
Most reports concentrate in \textit{Self-Action} (92.31\%), followed by \textit{Perception} (3.14\%), \textit{Agent Initialization} (3.09\%), \textit{Mutual Interaction} (1.09\%), and \textit{Evolution} (0.37\%). 
This distribution captures a coarse-grained manifestation pattern rather than normalized failure risk across lifecycle stages. 
\textit{Self-Action} covers the central execution loop of agent systems, including reasoning, planning, tool invocation, memory access, state processing, and task execution. 
Therefore, users may observe a failure in \textit{Self-Action} even when configuration, perception, dependency, or model output conditions trigger it. 
The lifecycle analysis thus shows where users and developers most often observe framework issues and highlights the need to further analyze the broad \textit{Self-Action} stage.\looseness=-1

\finding{
Reported failures concentrate in the broad \textit{Self-Action} stage.
This concentration does not indicate that \textit{Self-Action} has the highest normalized failure risk.
Instead, it shows that many framework issues become visible in the central execution loop.
This result motivates finer grained analysis of planning, tool invocation, memory and state handling, output parsing, execution control, and task result validation.
}

\subsection{RQ2: Functionality Challenges Associations}
\label{ssec:rq2}

To understand recurring patterns in the full filtered bug dataset, this section analyzes associations among reported causes, symptoms, and lifecycle stages. 
Following~\cite{haberman1973residuals,pearson1900criterion}, we use adjusted standardized residuals ($d_{ij}$) and chi square cell contributions ($\chi^2_{ij}$). 
The residual $d_{ij}$ captures the direction and strength of a cell-level deviation from independence, while $\chi^2_{ij}$ measures its contribution to the overall association. 
We regard a cell-level association as significant when its corresponding test satisfies $p \le 0.001$. We examine reported cause-symptom associations and lifecycle signatures, with heatmap visualizations available on our artifact site~\cite{our_website}.\looseness=-1

\noindent \textbf{Association between Root Cause and Symptom.}
The association matrix shows a sparse and concentrated structure, where a small number of reported causes account for many dominant symptom manifestations. \textit{Incorrect Functionality} and \textit{Crash} strongly associate with \textit{Misconfiguration} and \textit{Environment Incompatibility}, suggesting that agent framework correctness depends on consistent runtime environments, dependency versions, external services, and developer supplied configurations. \textit{Build Failure} strongly associates with \textit{Resource Limitation} ($d_{ij}=18.65$), \textit{Dependent Module Issue}, and \textit{Concurrency Issue}, indicating that build failures often relate to infrastructure constraints rather than only local implementation errors. 
In contrast, frequent API level causes, including \textit{API Incompatibility} and \textit{API Misuse}, scatter across multiple symptoms, suggesting that API issues act as cross cutting fault sources and require tests for API sequences, state transitions, and tool call contracts. 
Cross framework patterns further reveal architectural effects: \textit{Build Failure} and \textit{Misconfiguration} show strong associations across four frameworks except MetaGPT. 
And in \textit{LangChain}, \textit{Display Anomaly} strongly associates with \textit{Documentation Desync}, and \textit{Resource Limitation} strongly associates with \textit{Poor Performance}, likely reflecting its modular design, rapid API evolution, chain style orchestration, and intensive LLM calls.\looseness=-1

\noindent \textbf{Association across Lifecycle Stages.}
Lifecycle associations reveal stage-specific bug signatures. 
In \textit{Agent Initialization}, \textit{Documentation Desync} ($d_{ij}=10.80$, $\chi^2_{ij}=108.27$) and \textit{Misconfiguration} ($d_{ij}=10.37$, $\chi^2_{ij}=86.01$) show the strongest associations, reflecting environment preparation, dependency loading, and configuration issues. 
In \textit{Perception}, \textit{Misconfiguration} and \textit{Frontend Backend Mismatch} show strong associations ($d_{ij}=7.38$ and $3.59$), suggesting input ingestion, interface consistency, and preprocessing issues. 
In \textit{Mutual Interaction}, \textit{Concurrency Issue} ($d_{ij}=7.33$) likely reflects asynchronous conflicts in multi agent message passing and state synchronization, while \textit{Display Anomaly} in \textit{Evolution} ($d_{ij}=7.23$) suggests inconsistent interface states or erroneous feedback during reflection and state update. 
For \textit{Self-Action}, associations appear more diffuse because this stage contains most reported bugs and covers reasoning, planning, tool invocation, memory access, state processing, and execution. 

This pattern does not imply that every bug type inherently belongs to \textit{Self-Action}; rather, the central execution loop aggregates diverse fault sources and symptoms, making this stage a broad localization point for finer grained testing and diagnosis.\looseness=-1

\finding{
Reported cause and symptom associations are sparse, while lifecycle stages show distinct bug signatures.
These patterns call for association-aware and stage-aware testing, including configuration validation in \textit{Agent Initialization}, input checks in \textit{Perception}, concurrency testing in \textit{Mutual Interaction}, and trace-based validation in \textit{Self-Action}.\looseness=-1
}

\subsection{RQ3: Agent Framework Developer Evolution Requests}
\label{ssec:rq3}


\begin{table*}[!htbp]
\centering
\scriptsize
\renewcommand{\arraystretch}{1.15}
\caption{The Composition of Motivations.}
\label{tab:rq3_motivation_categories}
\begin{tabular}{|p{2.0cm}|p{1.3cm}|p{13.5cm}|}
\hline
\textbf{Category} & \textbf{Percentage} & \textbf{Explanation} \\
\hline
\textit{M1.Model Adaptation} & 15.70\% & The motivation is driven by differences among LLM vendors. The goal is to make the framework compatible with more models.\\
\hline
\textit{M2.Output Constraints} & 8.53\% & The motivation targets the form of model outputs. The goal is to make outputs more parseable, verifiable, and reusable, thereby reducing downstream parsing cost and improving stability. \\
\hline
\textit{M3.Tool Ecosystem} & 10.38\% & The  motivation targets the tool system itself. The goal is to expand callable capabilities or improve tool mechanisms. \\
\hline
\textit{M4.Retrieval Augmentation} & 3.21\% & The motivation targets the RAG pipeline. The goal is to improve knowledge acquisition and support more vector stores or retrieval backends. \\
\hline
\textit{M5.Memory Persistence} & 6.80\% & The motivation targets memory or state persistence. The goal is to reuse context across steps, tasks, or runs, enabling continuity beyond a single execution.\\
\hline
\textit{M6.Orchestration Expressiveness} & 18.54\% & The motivation targets workflow expression. The goal is to organize agents, tasks, and flows more flexibly. \\
\hline
\textit{M7.Runtime Efficiency} & 4.33\% & The motivation targets throughput, latency, or cost. The goal is to make execution faster and cheaper.\\
\hline
\textit{M8.Reliability and Security} & 3.09\% & The motivation targets operational reliability or compliance. The goal is to reduce failures, improve usability in enterprise network environments, and lower security risks. \\
\hline
\textit{M9.Observable Diagnosis} & 8.90\% & The  motivation targets explainability and diagnostics. The goal is to make internal execution visible, enabling faster troubleshooting and analysis.\\
\hline
\textit{M10.Development Delivery} & 16.81\% & The motivation targets engineering enablement. The goal is to reduce onboarding and integration cost, and improve delivery efficiency. \\
\hline
\textit{M11. Others} & 3.71\% &
The motivation in agent frameworks encompasses issues that do not fit into any of the previous ten categories. \\
\hline
\end{tabular}
\end{table*}


\begin{table*}[!htbp]
\centering
\scriptsize
\renewcommand{\arraystretch}{1.15}
\caption{The Composition of Requirements.}
\label{tab:rq3_requirement_categories}
\begin{tabular}{|p{2.5cm}|p{1.3cm}|p{13.0cm}|}
\hline
\textbf{Category} & \textbf{Percentage} & \textbf{Explanation} \\
\hline
\textit{R1.Integration Requirement} & 19.28\% & The requirement aims to integrate external tools, services, model providers, or third-party libraries with the agent framework to extend its capabilities. \\
\hline
\textit{R2.Feature Proposal} & 22.25\% & The  requirement calls for introducing entirely new features, tools, integrations, or capabilities, thereby extending the functional boundaries of the agent framework. \\
\hline
\textit{R3.Feature Enhancement} & 49.07\% & The  requirement focuses on optimizing, strengthening, or adjusting existing functionality to improve performance, usability, flexibility, or overall user experience. \\
\hline
\textit{R4.Documentation Improvement} & 7.05\% & The requirement requests updates, additions, or refinements to documentation. \\
\hline
\textit{R5.Infrastructure Optimization} & 1.73\% & The requirement concerns improvements to foundational support systems, including build processes, dependency management, packaging, deployment, and environment configuration. \\
\hline
\textit{R6.Others} & 0.62\% & The requirement in agent frameworks encompasses issues that do not fit into any of the previous five categories.\\
\hline

\end{tabular}
\end{table*}
To understand developer-facing requests in agent frameworks, we analyze 809 feature requests by lifecycle stage, motivation, and requirement. 
\autoref{tab:rq3_motivation_categories} and \autoref{tab:rq3_requirement_categories} report the 11 motivation categories and six requirement categories. This analysis identifies where developer demands concentrate and how they map to the agent lifecycle.

\noindent
\textbf{Analysis of Motivation Distribution.}
\autoref{tab:rq3_motivation_categories} shows that four motivations dominate feature requests: \textit{Orchestration Expressiveness} (18.54\%), \textit{Development Delivery} (16.81\%), \textit{Model Adaptation} (15.70\%), and \textit{Tool Ecosystem} (10.38\%), together accounting for 61.43\% of all requests. 
These motivations show that developers use agent frameworks as infrastructure beyond simple LLM wrappers and request support for workflow control, production delivery, model compatibility, and tool integration. 
Specifically, \textit{Orchestration Expressiveness} concerns conditional branches, dynamic routing, and long-horizon coordination; \textit{Development Delivery} concerns packaging, deployment, debugging, and engineering workflows; \textit{Model Adaptation} captures demand for diverse closed-source and open-source LLMs; and \textit{Tool Ecosystem} highlights the role of external APIs and tools in agent execution~\cite{schick2023toolformer,qin2024toolllm}. 
Across frameworks, these four motivations occupy leading positions, while \textit{Retrieval Augmentation} and \textit{Reliability and Security} appear less frequently, suggesting that reported requests focus more on foundational operational capabilities than on specialized extensions.\looseness=-1

\finding{
Feature request motivations concentrate on \textit{Orchestration Expressiveness}, \textit{Development Delivery}, \textit{Model Adaptation}, and \textit{Tool Ecosystem}.
This pattern shows a demand profile centered on orchestratable, deployable, adaptable, and integrable agent frameworks.
}

\noindent
\textbf{Analysis of Requirement Distribution.}
\autoref{tab:rq3_requirement_categories} shows that \textit{Feature Enhancement} dominates the requirement distribution (49.07\%), followed by \textit{Feature Proposal} (22.25\%) and \textit{Integration Requirement} (19.28\%), indicating stronger demand for refining existing capabilities than adding entirely new ones. 
This refinement-oriented demand profile emphasizes robustness, flexibility, usability, and convenience. 
Framework design also shapes requirement patterns: \textit{LangChain} and \textit{CrewAI} report frequent \textit{Integration Requirement}, reflecting their role as interoperability hubs for tools, services, and third-party components~\cite{langchainframework,wang2024survey}, while \textit{LangGraph} and \textit{MetaGPT} show more prominent \textit{Feature Enhancement}, suggesting stronger demand for mature intrinsic capabilities in graph-based orchestration and role-based collaboration.\looseness=-1

\finding{
The dominance of \textit{Feature Enhancement} (49.07\%) over \textit{Feature Proposal} (22.25\%) shows a refinement oriented demand profile.
Maintainers should prioritize robustness, flexibility, and developer convenience in existing features.
}

\noindent
\textbf{Analysis of Lifecycle Distribution.}
Mapping feature requests to the agent lifecycle shows that developer demands concentrate on \textit{Self-Action} (64.52\%), with \textit{Mutual Interaction} (17.31\%), \textit{Agent Initialization} (8.53\%), \textit{Evolution} (5.07\%), and \textit{Perception} (4.57\%) accounting for the rest. 
This distribution indicates where developers localize requested improvements, rather than normalized demand rates across lifecycle stages. 
\textit{Self-Action} ranks first because it covers the central execution loop, including reasoning, tool invocation, memory handling, state processing, and task execution. 
Compared with functionality challenges in~\autoref{sec:rq1}, feature requests cover more lifecycle stages: \textit{Mutual Interaction} increases from 1.09\% in bug reports to 17.31\% in feature requests, and \textit{Agent Initialization} increases from 3.09\% to 8.53\%.
This divergence suggests that bugs often surface in the execution loop, while developer requests also target coordination, initialization, and long term adaptation, reflecting demand for correct execution, smoother setup, richer collaboration, and persistent agent behavior.\looseness=-1

\finding{
Feature requests concentrate on \textit{Self-Action} (64.52\%) but cover more lifecycle stages than functionality challenges.
This pattern suggests that framework maintenance should strengthen the central execution loop while also addressing multi-agent coordination, initialization support, and long term adaptation.
}

\subsection{RQ4: Usability Concerns Associations}
\label{ssec:rq4}

To examine how developer facing requests interrelate, we analyze associations among motivations, requirements, and lifecycle stages. 
Following~\autoref{ssec:rq2}, we use adjusted standardized residuals ($d_{ij}$), chi-square cell contributions ($\chi^2_{ij}$), and the Haldane and Anscombe corrected log odds ratio ($\log \mathrm{OR}(\mathrm{HA})$)~\cite{weber2020zero,chen2010big}.
The first two metrics characterize cell-level deviations and contributions, while $\log \mathrm{OR}(\mathrm{HA})$ helps interpret sparse cells through a 0.5 correction to reconstructed 2$\times$2 tables. 
We examine motivation-requirement associations, framework-level differences, and lifecycle signatures.
Our artifact site provides heatmap visualizations~\cite{our_website}.\looseness=-1

\noindent
\noindent
\textbf{Association between Motivation and Requirement.}
The motivation and requirement association matrix reveals structured request patterns. 
\textit{Development Delivery} strongly associates with \textit{Documentation Improvement} ($d_{ij}=13.01$, $\chi^2_{ij}=130.91$, $\log \mathrm{OR}(\mathrm{HA})=3.27$) and \textit{Infrastructure Optimization} ($d_{ij}=7.68$), suggesting demand for clearer guides and stable infrastructure. 
\textit{Model Adaptation} concentrates on \textit{Integration Requirement} ($d_{ij}=10.9$), reflecting demand for connections to diverse model providers and vendor APIs.
Framework level results show consistent design-related patterns: in \textit{AutoGen}, \textit{Model Adaptation} strongly associates with \textit{Integration Requirement} ($d_{ij}=8.26$), aligning with dialogue-centered multi agent collaboration; 
in \textit{CrewAI}, \textit{Development Delivery} strongly associates with \textit{Documentation Improvement} ($d_{ij}=9.75$), suggesting stronger demand for engineering guidance.\looseness=-1

\noindent
\textbf{Association across Lifecycle Stages.}
Mapping motivations and requirements onto lifecycle stages reveals stage-specific request signatures. 
In \textit{Agent Initialization}, \textit{Documentation Improvement} shows a strong association ($d_{ij}=6.95$, $\chi^2_{ij}=41.12$, $\log \mathrm{OR}(\mathrm{HA})=1.95$), suggesting demand for configuration guides and quickstart materials during environment, credential, dependency, and initial agent setup. 
Other stage associations provide useful but more tentative signals because some stage-level cells contain limited samples.
\textit{Evolution} associates with \textit{Development Delivery} ($d_{ij}=4.33$, $\chi^2_{ij}=14.82$, $\log \mathrm{OR}(\mathrm{HA})=1.36$), suggesting that requests tend to emphasize delivery efficiency and engineering continuity during migration, updates, and long-term adaptation.\looseness=-1

\finding{
Developer facing requests exhibit structured associations among motivations, requirements, and lifecycle stages.
\textit{Development Delivery} aligns with documentation and infrastructure needs, while \textit{Model Adaptation} aligns with integration needs.
These patterns suggest that framework maintenance should organize feature requests by motivation, requirement type, and lifecycle context rather than treating them as isolated backlog items.\looseness=-1
}

\section{Discussion}
\label{sec:discuss}

\subsection{Implications}

Our findings show that agent framework quality assurance should jointly consider failure causes, symptoms, developer requests, and lifecycle stages. Findings 1 and 2 reveal semantic interface faults and silent functional failures. Findings 3, 4, and 7 highlight execution centered concentration and stage specific patterns. Findings 5, 6 and 8 show that developer requests offer structured signals for framework evolution. These results suggest the following implications.\looseness=-1

\textbf{Framework Maintainers.}
Findings 1 and 4 show that API, configuration, parsing, and serialization causes dominate many failures and vary across lifecycle stages. Maintainers should prioritize stable interface contracts, clearer configuration diagnostics, provider compatibility checks, and robust handling of states and structured outputs. Findings 5 and 6 further suggest that requests for orchestration expressiveness, development delivery, model adaptation, tool ecosystem support, and feature enhancement should guide framework evolution.\looseness=-1

\textbf{Framework Users.}
Findings 1 and 2 indicate that many failures occur at API, configuration, schema, state, and output boundaries, and often appear as incorrect or missing results rather than crashes. Users should check API versions, provider settings, tool schemas, serialized states, and execution traces, and validate tool calls, state transitions, structured outputs, and final results.\looseness=-1

\textbf{Testing Researchers.}
Findings 1, 2, and 4 suggest testing beyond crash centered and tensor level techniques. API issues motivate version aware API sequence testing; parsing errors motivate mutation of \texttt{JSON} objects, tool call payloads, and schema constrained responses; serialization errors motivate round trip state testing; incorrect functionality motivates execution trace replay and semantic output validation; and misconfiguration motivates configuration mutation with diagnostic and graceful failure oracles.\looseness=-1

\textbf{Broader Community.}
Findings 3 and 7 show that bugs and feature requests concentrate in the broad \textit{Self-Action} stage, while feature requests span more stages. Benchmarks, issue templates, and diagnostic tools should collect finer grained execution information, including planning, tool invocation, memory access, state transition, structured output parsing, and result validation, to improve triage, testing, debugging, and maintenance.\looseness=-1

\subsection{Limitations and Future Work}

Our study has several limitations that guide future work. We group them into two aspects: issue based empirical data and taxonomy construction with lifecycle modeling.\looseness=-1

First, our results depend on public GitHub issues from five widely used open source frameworks. GitHub labels and issue types guide initial filtering, and manual inspection removes false positives and low information reports. However, inconsistent labeling may omit unlabeled or mislabeled issues. Issue reports do not capture true bug incidence, since user exposure, reporting activity, and symptom visibility may inflate frequent categories, while private deployments and hard to diagnose failures may hide severe issues. Feature requests serve as proxies for developer facing improvement needs, but they may reflect usability friction, desired capabilities, ecosystem integration, or product direction. Long tail, domain specific, closed source, and enterprise frameworks may also follow different practices. Thus, our results characterize reported functionality challenges and developer requests rather than normalized failure rates or the full agent infrastructure ecosystem. Future work can incorporate keyword search, patch analysis, maintainer validation, telemetry, industrial bug trackers, deployment logs, surveys, interviews, and broader framework populations.\looseness=-1

Second, our analysis relies on expert judgment in taxonomy construction, issue labeling, and lifecycle mapping. We assign one primary category and one primary lifecycle stage to each issue to improve comparability, but this choice may simplify reports that involve multiple causes, symptoms, motivations, requirements, or lifecycle stages. In particular, \textit{Self-Action} marks where issues appear or where requests target, not necessarily where their causes originate. We mitigate annotation threats through multiple experienced annotators, randomized annotation order, explicit coding rules, consensus resolution, and Fleiss' kappa reporting. Future work can use multi label annotation, maintainer adjudication, model assisted coding, trace based validation, and finer grained execution stages, such as reasoning, planning, tool invocation, memory handling, state update, output parsing, and result validation.\looseness=-1

\section{Conclusion}
\label{sec:conclusion}

We investigate functionality challenges and usability concerns in modern agent frameworks. From 15,822 GitHub issues across five mainstream frameworks, we identify 5,669 bug reports and 809 feature requests, and construct a four-dimensional taxonomy aligned with a five-stage lifecycle. This mapping captures manifestation stages and operational targets rather than causal origins, so the concentration in \textit{Self-Action} reflects execution-centered reporting. Our findings highlight execution-centered failures and developer demand for refining existing capabilities, offering guidance for testing, debugging, maintenance, and evolution of reliable agent infrastructure.
\looseness=-1

\section{Data Availability}
The executable experiment code and all the results can be found on our website~\cite{our_website}.

{\scriptsize
  \bibliographystyle{IEEEtran}
  \bibliography{ref}
}

\end{document}